\documentclass[prd,aps,preprint,showpacs,nofootinbib,superscriptaddress]{revtex4}
\usepackage{epsfig}
\usepackage{color}
\usepackage{amsmath}
\newcommand{\ben}{\begin{eqnarray}}
\newcommand{\een}{\end{eqnarray}}
\newcommand{\nnu}{\nonumber\\}
\newcommand{\bef}{\begin{figure}[htb]\centering}
\newcommand{\eef}{\end{figure}}
\newcommand{\bet}{\begin{table}[hbt]\centering}
\newcommand{\eet}{\end{table}}

\newcommand{\shat}{\hat{s}}
\newcommand{\that}{\hat{t}}
\newcommand{\uhat}{\hat{u}}
\newcommand{\qbar}{\bar{q}}
\newcommand{\sla}[1]{{#1}\!\!\!\!\slash}

\begin{document}
\title{Dihadron Azimuthal Correlation from Collins Effect in Unpolarized Hadron Collisions}
\author{Zhong-Bo Kang}
\email{zkang@bnl.gov}
\affiliation{RIKEN BNL Research Center,
                Brookhaven National Laboratory,
                Upton, NY 11973, USA}
\author{Feng Yuan}
\email{fyuan@lbl.gov}
\affiliation{RIKEN BNL Research Center,
                Brookhaven National Laboratory,
                Upton, NY 11973, USA}
\affiliation{Nuclear Science Division,
                Lawrence Berkeley National Laboratory,
                Berkeley, CA 94720, USA}

\begin{abstract}
We study the dihadron azimuthal correlation produced
nearly back-to-back in unpolarized hadron collisions,
arising from the product of two Collins fragmentation
functions. Using the latest Collins fragmentation
functions extracted from the global analysis of available
experimental data, we make predictions for the azimuthal
correlation of two-pion production in $pp$ collisions at
RHIC energies. We find that the correlation is sizable in
the mid-rapidity region for moderate jet transverse momentum.
\end{abstract}
\date{\today}

\maketitle

%%%%%%%%%%%%%%
\section{Introduction}
The transverse momentum dependent (TMD) distributions and
fragmentation functions have received much attention
recently~\cite{D'Alesio:2007jt}. They are believed to be
responsible for several azimuthal asymmetries observed in
experiments, such as the single transverse spin asymmetry
(SSA) in semi-inclusive hadron production in
deep inelastic scattering (SIDIS)~\cite{Airapetian:2004tw, Ageev:2006da}
and in hadronic collisions~\cite{:2008qb},
as well as the large $\cos(2\phi)$ anomalous azimuthal
asymmetry in back-to-back dihadron production
in $e^+e^-$ annihilation~\cite{Seidl:2008xc}.

Among these TMD parton distributions and fragmentation functions,
the Sivers quark distribution~\cite{Siv90} and the Collins
fragmentation function~\cite{Collins93} are mostly discussed
in the last few years. The Sivers quark distribution
represents a distribution of unpolarized quarks in a
transversely polarized nucleon, through a correlation
between the quark's transverse momentum and the nucleon
polarization vector. On the other hand, the Collins
fragmentation function describes a transversely polarized
quark jet fragmenting into an unpolarized hadron, whose
transverse momentum relative to the jet axis correlates
with the transverse polarization vector of the fragmenting quark.

Though both of them belong to the so-called ``naive-time-reversal-odd''
(T-odd) functions, they have very different universality properties.
For the Sivers functions, it has been shown that they differ
by a sign for the SIDIS and Drell-Yan (DY) processes~\cite{Collins:2002kn},
and those in the hadronic collisions have even
more nontrivial relation to that in SIDIS and DY
processes~\cite{Bacchetta:2005rm, gl, QVY, VY, CQ}.
On the other hand, the Collins fragmentation function
is universal between different processes, in the SIDIS,
$e^+e^-$ and hadronic collisions~\cite{metz, gamberg, Yuan:2009dw}.
The effect of the Collins fragmentation function has
been recently explored by one of us in the azimuthal
asymmetric distribution of hadrons inside a jet in
$p^\uparrow p$ collision \cite{Yuan:2007nd}. It is
demonstrated that the asymmetry is sizable at RHIC,
therefore, the experimental study of this process
could provide an important information on the
universality of the Collins fragmentation function.

Another difference between these two functions is that
the Collins fragmentation function is chiral-odd whereas the
Sivers function is chiral-even.
Because of its chiral-odd nature, the Collins effect can only be
observed when it is coupled to another chiral-odd distribution
or fragmentation function. In SIDIS, the chiral-odd quark
transversity~\cite{trans} can couple to the Collins fragmentation
function and leads to nonzero azimuthal SSA~\cite{Collins93}.
This SSA has been studied by the HERMES~\cite{Airapetian:2004tw} and
COMPASS~\cite{Ageev:2006da} collaborations,
and very interesting results on the Collins fragmentation
function have been found. In $e^+e^-$ annihilation process,
two Collins fragmentation functions couple to each other in
the back-to-back dihadron production, results into to a
$\cos(2\phi)$ azimuthal asymmetry~\cite{boer}. This anomalous
$\cos(2\phi)$ asymmetry has been measured by the BELLE
Collaboration~\cite{Seidl:2008xc}, and was found consistent
with the HERMES and COMPASS measurements on the Collins
fragmentation functions.
Recently a global analysis of these experimental data has
been performed and the Collins fragmentation functions
have been extracted~\cite{Anselmino:2007fs}.

In this paper we investigate the possibility of exploring the
Collins fragmentation function in unpolarized $pp$ collision
by studying the azimuthal correlation in back-to-back dihadron
production, following the same wisdom of dihadron production in
$e^+e^-$ annihilation. We show that the asymmetry is
proportional to the product of two Collins fragmentation
function, same as that in $e^+e^-$ annihilation. Using the
latest Collins fragmentation function extracted from the
global analysis of available data on SIDIS and $e^+e^-$
experiments, we estimate the asymmetry for dihadron production
at RHIC energy. We find that the azimuthal asymmetry is sizable
at mid-rapidity region for moderate jet transverse momentum.
We argue that this process shall provide addtional
important information on the
Collins fragmentation function and its universality properties.

The rest of the paper is organized as follows.
In Sec.~\ref{cal}, we derive the theoretical results
for the dihadron azimuthal correlation produced nearly
back-to-back in unpolarized hadron collision. In
Sec.~\ref{estimate}, we present our numerical
predictions for the azimuthal correlation in unpolarized
$pp$ collisions for RHIC kinematics. Finally, we summarize
our findings and the corresponding conclusions in Sec.~\ref{con}.

%%%%%%%%%%%%%%
\section{Dihadron azimuthal correlation in unpolarized hadron collision}
\label{cal}
We study the azimuthal correlation of two hadrons $h_1$
and $h_2$ produced nearly back-to-back in a hadronic collision,
\ben
A(P_1)+B(P_2)\to h_1(P_{h1})+h_2(P_{h2})+X,
\een
where both of the incident hadrons $A$ and $B$ are
unpolarized. The momenta of the initial hadrons are
denoted by $P_1$ and $P_2$, and those of the final
hadrons by $P_{h1}$ and $P_{h2}$, respectively.

%%%%%%%%%%%%%%%%%%%%%%%%%%%%%%%%%%%%%%%%%%
\bef
\psfig{file=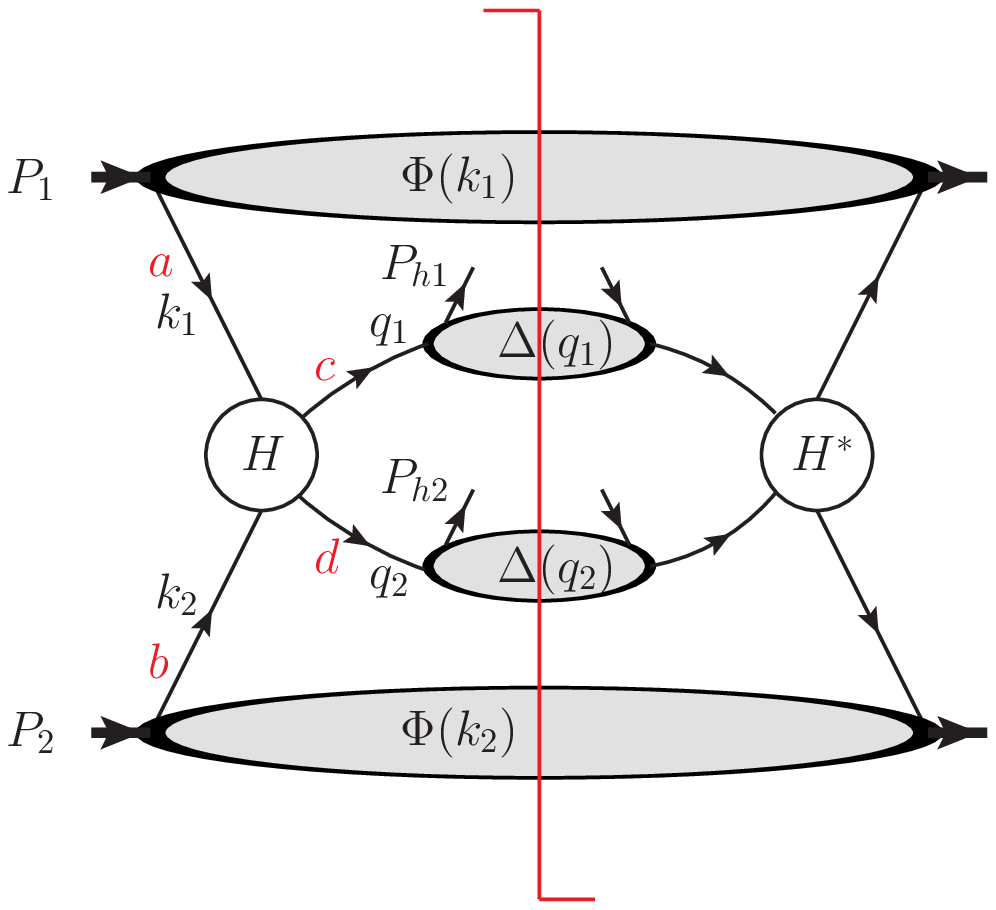, width=2.5in}
\caption{The leading order contribution to the cross
section of $A(P_1)+B(P_2)\to h_1(P_{h1})+h_2(P_{h2})+X$
with the $2\to 2$ partonic process
$a(k_1)+b(k_2)\to c(q_1)+d(q_2)$.}
\label{parton}
\eef
%%%%%%%%%%%%%%%%%%%%%%%%%%%%%%%%%%%%%%%%%%

The leading order contribution to the scattering
cross section comes from partonic $2\to 2$ sub-processes,
$a(k_1)+b(k_2)\to c(q_1)+d(q_2)$, as shown in Fig.~\ref{parton}.
The parton momenta are expanded as follows,
\begin{subequations}
\begin{align}
k_1&=x_1P_1+k_{1T},\\
k_2&=x_2P_2+k_{2T},\\
P_{h1}&=z_1q_1+p_{1T},\\
P_{h2}&=z_2q_2+p_{2T},
\end{align}
\end{subequations}
where $x_1$ and $x_2$ are the longitudinal momentum fractions,
and $k_{1T}$ and $k_{2T}$ are the transverse momentum of the
parton relative to the corresponding incident hadron. $q_1$
and $q_2$ are the momenta of the nearly back-to-back jets
$J_1$ and $J_2$, which has a polar angle $\theta_1$ and
$\theta_2$ relative to the incoming hadron $P_1$, respectively.
The momenta of the incoming hadrons and the final state two jets
form the so-called reaction plane (approximately).
Besides carrying a longitudinal momentum fraction
$z_1$ ($z_2$) of the jet $J_1$ ($J_2$), the hadron
$h_1$ ($h_2$) also has a transverse momentum
$p_{1T}$ ($p_{2T}$) relative to the jet $J_1$ ($J_2$)
direction, which defines an azimuthal angle with the
reaction plane: $\phi_1$ ($\phi_2$), as shown in
Fig.~\ref{show}. Due to the Collins effect, there
will be a correlation between these two azimuthal
angles $\phi_1$ and $\phi_2$, which is proportional
to the product of the two Collins fragmentation
functions as we will show below.

%%%%%%%%%%%%%%%%%%%%%%%%%%%%%%%%%%%%%%%%%%
\bef
\psfig{file=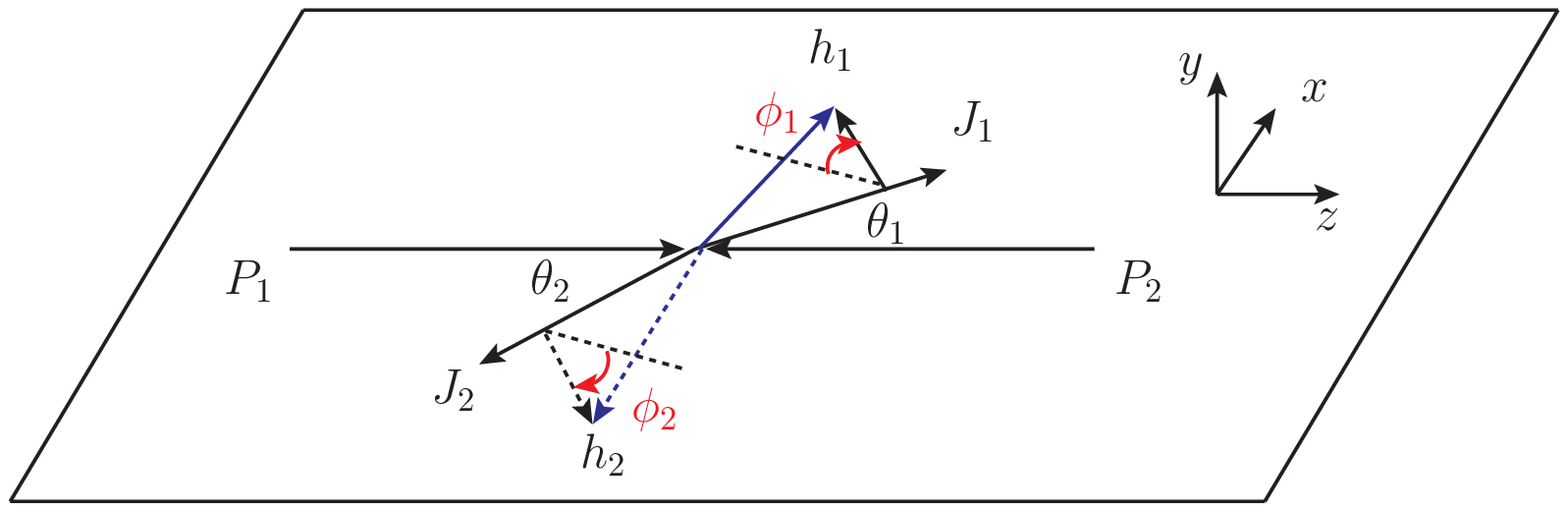, width=4.0in}
\caption{Illustration of the kinematics for dihadron production $AB\to h_1h_2+X$.}
\label{show}
\eef
%%%%%%%%%%%%%%%%%%%%%%%%%%%%%%%%%%%%%%%%%%

From Fig.~\ref{parton}, the cross section of dihadron production
can be written as~\cite{Bacchetta:2005rm},
\ben
d\sigma&=&\frac{1}{2S}\int dx_1 d^2k_{1T}dx_2 d^2k_{2T}dz_1 d^2p_{1T}dz_2 d^2p_{2T}
\frac{d^3q_1}{(2\pi)^3 2E_1} \frac{d^3q_2}{(2\pi)^3 2E_2}(2\pi)^4\delta^4(k_1+k_2-q_1-q_2)
\nnu
&&
\times
{\rm Tr}\left[\Phi(x_1, k_{1T})\Phi(x_2, k_{2T})\Delta(z_1, p_{1T})\Delta(z_2, p_{2T})H(k_1,k_2,q_1,q_2)H^*(k_1,k_2,q_1,q_2)\right],
\label{tot}
\een
where $S=(P_1+P_2)^2$, and $\Phi(x, k_{T})$ and $\Delta(z, p_{T})$
are the distribution and fragmentation correlation functions, and
$H(k_1, k_2, q_1, q_2)$ are the hard part amplitude.

For the unpolarized hadron, the correlation functions
$\Phi(x, k_{T})$ can be simply decomposed as~\cite{Mulders:1995dh, Boer:1997nt},
\ben
\Phi(P, x, k_{T})=\frac{1}{2}\left[f(z, k_T^2)\sla{P}+h_1^{\perp}(z, k_T^2)\frac{\sigma^{\mu\nu}k_{T\mu}P_{\nu}}{M}\right],
\een
where $f(x, k_T^2)$ is the unpolarized TMD parton distribution
function, and $h_1^{\perp}(z, k_T^2)$ is the Boer-Mulders
function~\cite{Boer:1997nt}. Similarly, we can parameterize
the gluon distributions from the incoming hadrons. The effect of
Boer-Mulders function in unpolarized hadronic collisions has
been extensively studied previously~\cite{Boer:2002ju, Boer:2007nd, Lu:2008qu},
which will be neglected in our current study.
We will concentrate on the effect coming from
the fragmentation correlation function $\Delta(z, p_{T})$,
which can be expanded as~\cite{Bacchetta:2006tn}
\ben
\Delta(P_h, z, p_T)=\frac{1}{2}\left[D(z, p_T^2)\sla{P}_h+H_1^{\perp}(z, p_T^2)\frac{\sigma^{\mu\nu}P_{h\mu}p_{T\nu}}{zM_h}\right],
\een
where $D(z, p_T^2)$ is the unpolarized TMD fragmentation function,
and $H_1^{\perp}$ is the Collins function. From its definition, we can
see that the Collins function describes a transversely polarized
quark jet fragmenting into an unpolarized hadron~\cite{Collins93}.
It is this Collins function that generates a non-vanishing
azimuthal correlation between the final state two hadrons.
Since Collins function is a chiral-odd TMD function,
the azimuthal correlation will depend on the product
of two Collins functions.

The phase space integral in Eq.~(\ref{tot}) can be simplified by using
\ben
&&\frac{d^3q_1}{(2\pi)^3 2E_1} \frac{d^3q_2}{(2\pi)^3 2E_2}(2\pi)^4\delta^4(k_1+k_2-q_1-q_2)
\nnu
&&=dy_1 dy_2 d P_\perp^2
\left(\frac{1}{8\pi S}\right)
\delta\left[x_1-\frac{P_\perp}{\sqrt{S}}\left(e^{y_1}+e^{y_2}\right)\right]
\delta\left[x_2-\frac{P_\perp}{\sqrt{S}}\left(e^{-y_1}+e^{-y_2}\right)\right],
\label{x1x2}
\een
where $y_1$ and $y_2$ are rapidities for the jet $J_1$ and
$J_2$, $P_\perp$ is the jet transverse momentum. Since
$p_{1T}, p_{2T}\ll P_\perp$, the rapidity of the hadron
is approximately equal to that of the parent jet. As
stated above, we neglect the intrinsic transverse momentum
effects for the incoming partons. Finally we obtain the
cross section for this process as
\ben
\frac{d\sigma}{d{\cal PS}}
&=&\frac{\pi\alpha_s^2}{S^2}\sum_{abcd}\frac{f_a(x_1)}{x_1}\frac{f_b(x_2)}{x_2}
\left[D_{c\to h_1}(z_1,p^2_{1T})D_{d\to h_2}(z_2,p^2_{2T})
H^U_{ab\to cd}\left(\shat,\that,\uhat\right)
\right.
\nnu
&~&
\left.
+\left(
p_{1T}\cdot p_{2T}-\frac{x_1x_2}{P_\perp^2}(P_1\cdot p_{1T} P_2\cdot p_{2T}+P_1\cdot p_{2T} P_2\cdot p_{1T})\right)
\nonumber\right.\\
&&\times \frac{H_1^\perp(z_1,p^2_{1T})}{z_1M_h}\frac{H_1^\perp(z_2,p^2_{2T})}{z_2M_h}
\left.H^{\rm Collins}_{ab\to cd}\left(\shat,\that,\uhat\right)
\right],
\label{main}
\een
where $d{\cal PS}\equiv dy_1 dy_2 dP_\perp^2 dz_1 dz_2 d^2p_{1T} d^2p_{2T}$
is the phase space for this process, $f_a(x_1)$ and $f_b(x_2)$
are the standard unpolarized parton distribution functions,
and $\shat$, $\that$, and $\uhat$ are the usual partonic
Mandelstam variables. The parton momentum fractions $x_1$
and $x_2$ are fixed by the delta functions in Eq.~(\ref{x1x2}),
\begin{subequations}
\begin{align}
x_1&=\frac{P_\perp}{\sqrt{S}}\left(e^{y_1}+e^{y_2}\right),
\\
x_2&=\frac{P_\perp}{\sqrt{S}}\left(e^{-y_1}+e^{-y_2}\right).
\end{align}
\end{subequations}

The normal partonic cross sections $H^U_{ab\to cd}$ are well-known \cite{Owens:1986mp},
\ben
H^U_{gg\to q\qbar}&=&\frac{1}{2N_c}\left[\frac{\that}{\uhat}+\frac{\uhat}{\that}\right]-\frac{N_c}{N_c^2-1}\left[\frac{\that^2+\uhat^2}{\shat^2}\right],
\\
H^U_{q\qbar\to q'\qbar'}&=&\frac{N_c^2-1}{2N_c^2}\left[\frac{\that^2+\uhat^2}{\shat^2}\right],
\\
H^U_{q\qbar\to q\qbar}&=&\frac{N_c^2-1}{2N_c^2}\left[\frac{\that^2+\uhat^2}{\shat^2}+\frac{\shat^2+\uhat^2}{\that^2}\right]-\frac{N_c^2-1}{N_c^3}\left[\frac{\uhat^2}{\shat\that}\right],
\\
H^U_{qq'\to qq'}&=&\frac{N_c^2-1}{2N_c^2}\left[\frac{\shat^2+\uhat^2}{\that^2}\right],
\\
H^U_{qq\to qq}&=&\frac{N_c^2-1}{2N_c^2}\left[\frac{\shat^2+\uhat^2}{\that^2}+\frac{\shat^2+\that^2}{\uhat^2}\right]-\frac{N_c^2-1}{N_c^3}\left[\frac{\shat^2}{\that\uhat}\right],
\\
H^U_{q\qbar\to gg}&=&\frac{4(N_c^2-1)}{N_c^3}\left[\frac{\that}{\uhat}+\frac{\uhat}{\that}\right]-\frac{N_c^2-1}{N_c}\left[\frac{\that^2+\uhat^2}{\shat^2}\right],
\\
H^U_{gq\to gq}&=&-\frac{N_c^2-1}{2N_c^2}\left[\frac{\shat}{\uhat}+\frac{\uhat}{\shat}\right]+\left[\frac{\shat^2+\uhat^2}{\that^2}\right],
\\
H^U_{gg\to gg}&=&\frac{4N_c^2}{N_c^2-1}\left[3-\frac{\that\uhat}{\shat^2}-\frac{\shat\uhat}{\that^2}-\frac{\shat\that}{\uhat^2}\right].
\een
The new hard parts $H^{\rm Collins}_{ab\to cd}$ that are
responsible for the azimuthal correlation are given by,
\ben
H^{\rm Collins}_{gg\to q\qbar}&=&\frac{1}{N_c}-\frac{N_c}{N_c^2-1}\left[\frac{2\that\uhat}{\shat^2}\right],
\\
H^{\rm Collins}_{q\qbar\to q'\qbar'}&=&\frac{N_c^2-1}{N_c^2}\left[\frac{\that\uhat}{\shat^2}\right],
\\
H^{\rm Collins}_{q\qbar\to q\qbar}&=&\frac{N_c^2-1}{N_c^2}\left[\frac{\that\uhat}{\shat^2}\right]-\frac{N_c^2-1}{N_c^3}\left[\frac{\uhat}{\shat}\right],
\\
H^{\rm Collins}_{qq'\to qq'}&=&0,
\\
H^{\rm Collins}_{qq\to qq}&=&\frac{N_c^2-1}{N_c^3},
\een
and the hard parts for the channels with gluon in the final state, $q\qbar\to gg$,
$qg\to qg$ and $gg\to gg$ vanish since there is no gluon Collins function.

%%%%%%%%%%%%%%
\section{Phenomenology study}
\label{estimate}
In this section we first properly define the azimuthal
asymmetry to be measured in the experiments. We then
use the latest Collins fragmentation function to
estimate this asymmetry for RHIC kinematics.

From Eq.~(\ref{main}), the explicit form of the azimuthal
correlation depends on the following function,
\ben
h(p_{1T}, p_{2T}, \phi_1, \phi_2)
\equiv
p_{1T}\cdot p_{2T}-\frac{x_1x_2}{P_\perp^2}(P_1\cdot p_{1T} P_2\cdot p_{2T}+P_1\cdot p_{2T} P_2\cdot p_{1T}).
\een
Function $h(p_{1T}, p_{2T}, \phi_1, \phi_2)$ is reduced to
its simplest form in the partonic center of mass frame,
\ben
h(p_{1T}, p_{2T}, \phi_1, \phi_2)=-p_{1T} p_{2T}\cos(\phi_1+\phi_2)\ .
\label{cmh}
\een
To take advantage of this simplicity, one could boost
from the Lab frame to the partonic CM frame experimentally.
Or equivalently, one can select the events with
$y_1+y_2\approx 0$, where the Lab frame
coincides with the partonic CM frame.
In the rest of our paper, we will take $y_1+y_2=0$ in our
calculations and present the numerical estimate for the
azimuthal asymmetry.

Integrating over the intrinsic transverse momentum $p_{1T}$ and $p_{2T}$, we have
\ben
\frac{d\sigma}{dy_1 dy_2 dP_\perp^2 dz_1 dz_2}=
\frac{\pi\alpha_s^2}{S^2}\sum_{abcd}\frac{f_a(x_1)}{x_1}\frac{f_b(x_2)}{x_2}
D_{c\to h_1}(z_1)D_{d\to h_2}(z_2)H^U_{ab\to cd}\left(\shat,\that,\uhat\right)
\een
While integrating over the moduli of the intrinsic momenta $p_{1T}$ and $p_{2T}$, and over the azimuthal angle $\phi_1$, one obtain
\ben
\frac{d\sigma}{dy_1 dy_2 dP_\perp^2 dz_1 dz_2d\left(\phi_1+\phi_2\right)}&=&
\frac{\alpha_s^2}{2 S^2}\sum_{abcd}\frac{f_a(x_1)}{x_1}\frac{f_b(x_2)}{x_2}
\left[D_{c\to h_1}(z_1)D_{d\to h_2}(z_2)H^U_{ab\to cd}\left(\shat,\that,\uhat\right)
\right.
\nnu
&&
\left.
-\cos(\phi_1+\phi_2)
\delta\hat{q}_{c\to h_1}^{(1/2)}(z_1)\delta\hat{q}_{d\to h_2}^{(1/2)}(z_2)
H^{\rm Collins}_{ab\to cd}\left(\shat,\that,\uhat\right)\right],
\een
where $\delta{\hat{q}}^{(1/2)}(z)$ is the so-called half-moment of the Collins function given by,
\ben
\delta{\hat{q}}^{(1/2)}(z)=\int d^2 p_{T}p_T \frac{H_1^\perp(z, p_T^2)}{zM_h}.
\een
Then following the normal analysis in $e^+e^-\to h_1h_2+X$, we define
\ben
A^{h_1h_2}&&(y_1, y_2, P_\perp, z_1, z_2, \phi_1+\phi_2)\nonumber\\
&&\equiv
\left.
\frac{d\sigma}{dy_1 dy_2 dP_\perp^2 dz_1 dz_2d\left(\phi_1+\phi_2\right)}
\right/
\frac{1}{2\pi}\frac{d\sigma}{dy_1 dy_2 dP_\perp^2 dz_1 dz_2}
\nnu
&&=1-\cos(\phi_1+\phi_2)
\frac{\sum_{abcd}f_a(x_1)f_b(x_2)
\delta\hat{q}_{c\to h_1}^{(1/2)}(z_1)\delta\hat{q}_{d\to h_2}^{(1/2)}(z_2)
H^{\rm Collins}_{ab\to cd}}
{\sum_{abcd}f_a(x_1)f_b(x_2)
D_{c\to h_1}(z_1)D_{d\to h_2}(z_2)H^U_{ab\to cd}}\ .
\een
%%%%%
Again to eliminate the false asymmetries~\cite{Seidl:2008xc},
we introduce the ratio of the unlike-sign to like-sign pion
production, $A_U$ and $A_L$, given by
\ben
R&\equiv& \frac{A_U}{A_L}=\frac{1-\cos(\phi_1+\phi_2)P_U}{1-\cos(\phi_1+\phi_2)P_L}
\approx
1-\cos(\phi_1+\phi_2)\left(P_U-P_L\right)
\nnu
&\equiv& 1-\cos(\phi_1+\phi_2) A_{12}(y_1, y_2, P_\perp, z_1, z_2)
\label{A12}
\een
with
\ben
&&P_U=\frac{\sum_{abcd}f_a(x_1)f_b(x_2)
\left[\delta\hat{q}_{c\to \pi^+}^{(1/2)}(z_1)\delta\hat{q}_{d\to \pi^-}^{(1/2)}(z_2)
+\delta\hat{q}_{c\to \pi^-}^{(1/2)}(z_1)\delta\hat{q}_{d\to \pi^+}^{(1/2)}(z_2)\right]
H^{\rm Collins}_{ab\to cd}}
{\sum_{abcd}f_a(x_1)f_b(x_2)
\left[D_{c\to \pi^+}(z_1)D_{d\to \pi^-}(z_2)+D_{c\to \pi^-}(z_1)D_{d\to \pi^+}(z_2)\right]
H^U_{ab\to cd}},
\\
&&P_L=\frac{\sum_{abcd}f_a(x_1)f_b(x_2)
\left[\delta\hat{q}_{c\to \pi^+}^{(1/2)}(z_1)\delta\hat{q}_{d\to \pi^+}^{(1/2)}(z_2)
+\delta\hat{q}_{c\to \pi^-}^{(1/2)}(z_1)\delta\hat{q}_{d\to \pi^-}^{(1/2)}(z_2)\right]
H^{\rm Collins}_{ab\to cd}}
{\sum_{abcd}f_a(x_1)f_b(x_2)
\left[D_{c\to \pi^+}(z_1)D_{d\to \pi^+}(z_2)+D_{c\to \pi^-}(z_1)D_{d\to \pi^-}(z_2)\right]
H^U_{ab\to cd}},
\\
&&A_{12}(y_1, y_2, P_\perp, z_1, z_2)=P_U-P_L.
\een

This way the true asymmetry due to Collins effect is
encoded in the so-called double ratio asymmetry
parameter $A_{12}(y_1, y_2, P_\perp, z_1, z_2)$.
To evaluate $A_{12}$ for the dihadron production
in unpolarized $pp$ collision at $\sqrt{S}=200$GeV
at RHIC, we use the Collins fragmentation functions~\cite{Anselmino:2007fs}
extracted from a combined fit to the experimental
data from HERMES, COMPASS and BELLE collaborations.
We use CTEQ5L parton distributions~\cite{Lai:1999wy}
and Kretzer unpolarized fragmentation function
obtained in~\cite{Kretzer:2000yf}.

%%%%%%%%%%%%%%%%%%%%%%%%%%%%%%%%%%%%%%%%%%
\bef
\psfig{file=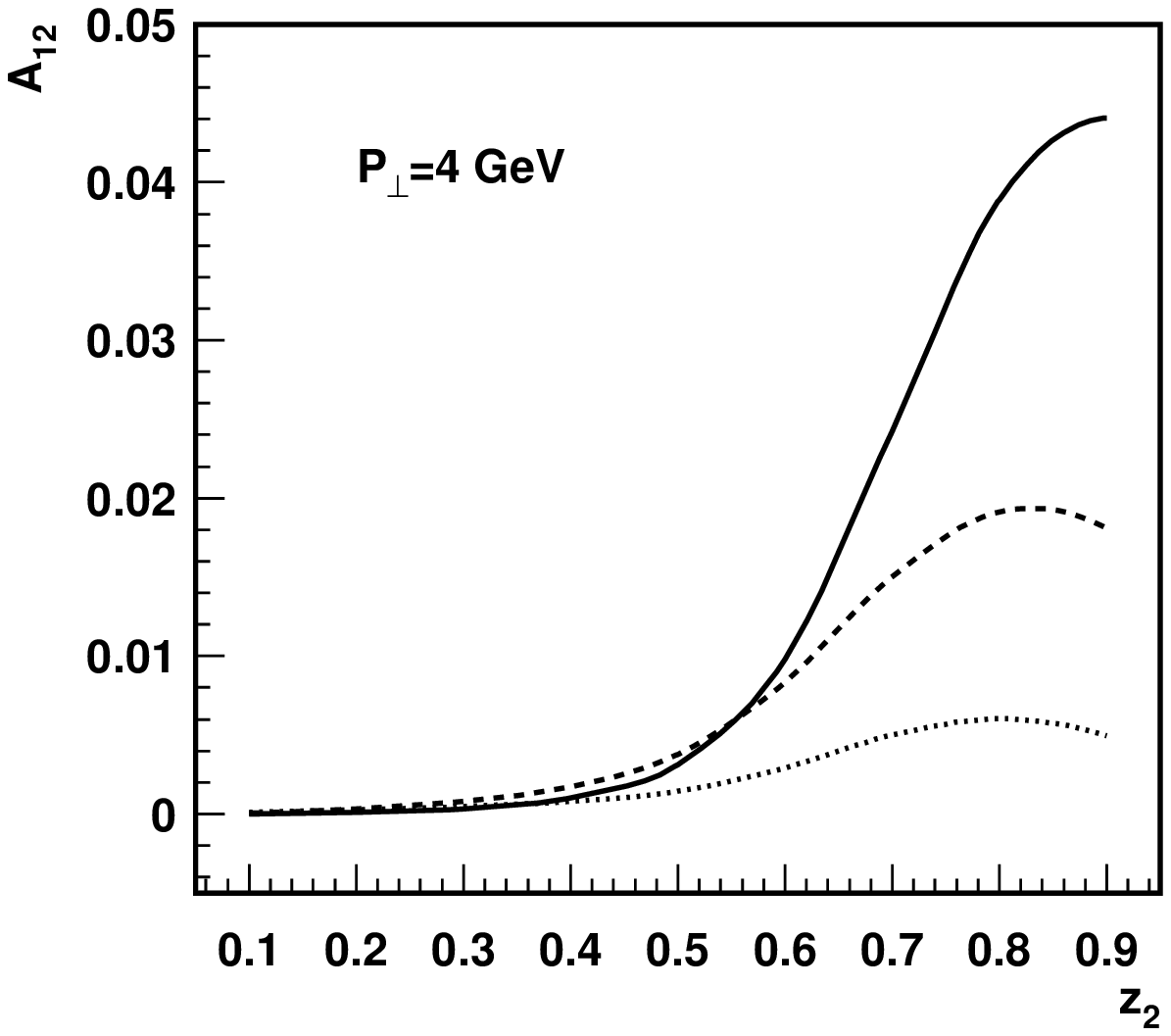, width=2.8in}\hskip 0.2in
\psfig{file=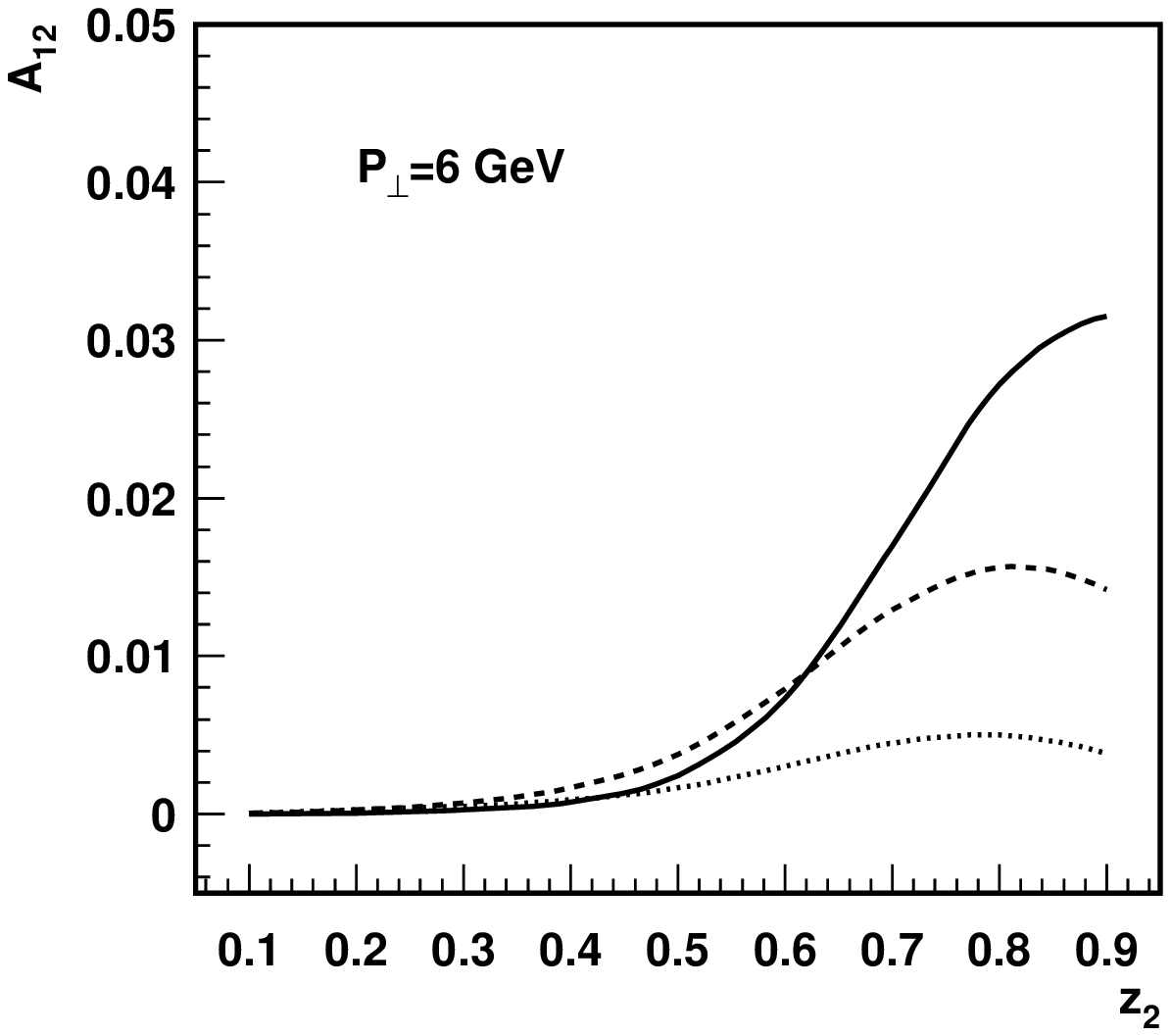, width=2.8in}
\caption{Azimuthal asymmetry ratios $A_{12}(y_1, y_2, P_\perp, z_1, z_2)$
(defined in Eq.~(\ref{A12})) of unlike-sign to like-sign pion
production in unpolarized proton-proton scattering at RHIC energy
$\sqrt{S}=200$GeV, as functions
of $z_2$ with $y_1=y_2=0$, $P_\perp=4$ GeV
(left) or $P_\perp=6$ GeV (right).
The curves are: solid ($0.7<z_1<0.9$),
dashed ($0.5<z_1<0.7$) and dotted ($0.3<z_1<0.5$).}
\label{plot}
\eef
%%%%%%%%%%%%%%%%%%%%%%%%%%%%%%%%%%%%%%%%%%

In Fig.~\ref{plot}, we plot $A_{12}(y_1, y_2, P_\perp, z_1, z_2)$
for the dihadron production in mid-rapidity $y_1=y_2=0$
region at RHIC energy $\sqrt{S}=200$ GeV. In the figure
on the left, we plot $A_{12}$ as a function of $z_2$ at
the jet transverse momentum $P_\perp=4$ GeV with $z_1$
integrated from three different ranges $0.3<z_1<0.5$ (dotted),
$0.5<z_1<0.7$ (dashed), and $0.7<z_1<0.9$ (solid), respectively.
On the right, we present the same plot but with $P_\perp=6$ GeV.
We find that the asymmetry $A_{12}$ is largest when
both $z_1$ and $z_2$ become large, same as what has
been observed in $e^+e^-$ experiments~\cite{Seidl:2008xc}.
On the other hand, $A_{12}$ decreases when increasing $P_\perp$.
This is also consistent with what BELLE observed if
one realizes that $\shat/4P_\perp^2=\sin^2\theta$ in
parton CM frame. Though it has similar features as that
in $e^+e^-$ collision, the asymmetry in hadronic collision
is smaller. This is due to the fact that there is copious
$gg\to gg$ and $qg\to qg$ contribution to the azimuthal
angle independent cross section, while they do not
contribute to the azimuthal dependent part since
there is no gluon Collins function. However, the
asymmetry is still around several percent and shall be
measurable at RHIC.

These results can be extended to general kinematics,
for example, in two different rapidity regions: $|y_1|\neq |y_2|$.
In this case, although the azimuthal angular dependence is not exactly 
as $\cos(\phi_1+\phi_2)$ in Eq.~(\ref{cmh}), the Collins fragmentation functions
will nevertheless lead to a nonzero mean value of 
$\langle \cos(\phi_1+\phi_2)\rangle$. This can be seen from 
the differential cross section expression in Eq.~(\ref{main}). 
The experimental observation of this nonzero effects can be used as 
signal of the Collins effects, since the 
normal fragmentation functions $D(z,p_{T})$ will not
contribute to a nonzero $\langle\cos(\phi_1+\phi_2)\rangle$. 
We hope that the future RHIC experiments can carry out these
measurements, and provide more information
on the Collins fragmentation functions, which will help us
to pin down the mechanism for the single spin asymmetry
in hadronic collisions as we discussed in the Introduction.

%%%%%%%%%%%%%%
\section{Summary}
\label{con}

In this paper, we have studied the dihadron azimuthal correlation
produced nearly back-to-back in unpolarized hadron
collision, arising from the product of two Collins
fragmentation functions. Using the latest Collins
fragmentation function extracted from the global
analysis of available experimental data, we make
predictions for the azimuthal correlation of two-pion
production in unpolarized $pp$ collisions at RHIC energies.
We find that the feature of the asymmetry is similar
to those observed in $e^+e^-$ annihilation. The asymmetry
parameter $A_{12}$ is sizable in the mid-rapidity region
for moderate jet transverse momentum, and could be measurable
in the experiments. The experimental study of this process
could provide the important information on the size of
Collins fragmentation function in hadronic collision, at
the same time, it could also be used to test the
universality properties of the Collins fragmentation
function in different processes.

\section*{Acknowledgments}

We thank M. Grosse Perdekamp, J. Qiu and R. Seidl for
helpful discussions.
This work was supported in part by the U. S. Department of Energy
under Grant No.~DE-FG02-05CH11231. We are
grateful to RIKEN, Brookhaven National Laboratory,
and the U.S. Department of Energy
(Contract No.~DE-AC02-98CH10886) for
providing the facilities essential
for the completion of this work.

%%%%%%%%%%%%%%


\begin{thebibliography}{99}

\bibitem{D'Alesio:2007jt}
For a review, see:
  U.~D'Alesio and F.~Murgia,
  %``Azimuthal and Single Spin Asymmetries in Hard Scattering Processes,''
  Prog.\ Part.\ Nucl.\ Phys.\  {\bf 61}, 394 (2008)
  [arXiv:0712.4328 [hep-ph]].

\bibitem{Airapetian:2004tw}
  A.~Airapetian {\it et al.}  [HERMES Collaboration],
  %``Single-spin asymmetries in semi-inclusive deep-inelastic scattering on  a
  %transversely polarized hydrogen target,''
  Phys.\ Rev.\ Lett.\  {\bf 94}, 012002 (2005)
  [arXiv:hep-ex/0408013].

\bibitem{Ageev:2006da}
  E.~S.~Ageev {\it et al.}  [COMPASS Collaboration],
  %``A new measurement of the Collins and Sivers asymmetries on a transversely
  %polarised deuteron target,''
  Nucl.\ Phys.\  B {\bf 765}, 31 (2007)
  [arXiv:hep-ex/0610068].

\bibitem{:2008qb}
  B.~I.~Abelev {\it et al.}  [STAR Collaboration],
%``Measurement of Transverse Single-Spin Asymmetries for Di-Jet Production
%in Proton-Proton Collisions at $\sqrt{s} = 200$ GeV,''
  Phys.\ Rev.\ Lett.\  {\bf 99}, 142003 (2007)
  [arXiv:0705.4629 [hep-ex]].

\bibitem{Seidl:2008xc}
  R.~Seidl {\it et al.}  [Belle Collaboration],
  %``Measurement of azimuthal asymmetries in inclusive production of hadron
  %pairs in e+ e- annihilation at Belle,''
  Phys.\ Rev.\ Lett.\  {\bf 96}, 232002 (2006)
  [arXiv:hep-ex/0507063];
  %``Measurement of Azimuthal Asymmetries in Inclusive Production of Hadron
  %Pairs in e+e- Annihilation at \sqrt{s} = 10.58 GeV,''
  Phys.\ Rev.\  D {\bf 78}, 032011 (2008)
  [arXiv:0805.2975 [hep-ex]].

\bibitem{Siv90}
D.~W.~Sivers,
%``Single Spin Production Asymmetries From The Hard Scattering Of
% Point- Like
%Constituents,''
Phys.\ Rev.\ D {\bf 41}, 83 (1990);
%``Hard Scattering Scaling Laws For Single Spin Production
% Asymmetries,''
Phys.\ Rev.\ D {\bf 43}, 261 (1991).

\bibitem{Collins93}
  J.~C.~Collins,
  %``Fragmentation of transversely polarized quarks probed in transverse
  %momentum distributions,''
  Nucl.\ Phys.\  B {\bf 396}, 161 (1993).

\bibitem{Collins:2002kn}
  J.~C.~Collins,
  %``Leading-twist Single-transverse-spin asymmetries: Drell-Yan and
  %Deep-Inelastic Scattering,''
  Phys.\ Lett.\  B {\bf 536}, 43 (2002).
  D.~Boer {\it et al.}, %P.~J.~Mulders and F.~Pijlman,
  %``Universality of T-odd effects in single spin and azimuthal asymmetries,''
  Nucl.\ Phys.\  B {\bf 667}, 201 (2003);
  Z.~B.~Kang and J.~W.~Qiu,
  %``Testing the Time-Reversal Modified Universality of the Sivers Function,''
  Phys.\ Rev.\ Lett.\  {\bf 103}, 172001 (2009).

\bibitem{gl}
C.~J.~Bomhof, P.~J.~Mulders and F.~Pijlman,
  %``Gauge link structure in quark quark correlators in hard processes,''
  Phys.\ Lett.\ B {\bf 596}, 277 (2004);
  Eur.\ Phys.\ J.\ C {\bf 47}, 147 (2006);
  JHEP {\bf 0702}, 029 (2007);
  C.~J.~Bomhof and P.~J.~Mulders,
  %``Non-universality of transverse momentum dependent parton distribution
  %functions,''
  Nucl.\ Phys.\  B {\bf 795}, 409 (2008)
  [arXiv:0709.1390 [hep-ph]].

\bibitem{QVY}
J.~W.~Qiu, W.~Vogelsang and F.~Yuan,
  %``Asymmetric Di-jet Production in Polarized Hadronic Collisions,''
  Phys.\ Lett.\  B {\bf 650}, 373 (2007);
  %``Single Transverse-Spin Asymmetry in Hadronic Dijet Production,''
  Phys.\ Rev.\ D {\bf 76}, 074029 (2007).

\bibitem{VY}
  W.~Vogelsang and F.~Yuan,
  %``Hadronic Dijet Imbalance and Transverse-Momentum Dependent Parton
  %Distributions,''
  Phys.\ Rev.\  D {\bf 76}, 094013 (2007).

\bibitem{CQ}
J.~Collins and J.~W.~Qiu,
  %``k_T factorization is violated in production of high-transverse-momentum
  %particles in hadron-hadron collisions,''
  Phys.\ Rev.\  D {\bf 75}, 114014 (2007);
  J.~Collins,
  %``2-soft-gluon exchange and factorization breaking,''
  arXiv:0708.4410 [hep-ph].

\bibitem{Bacchetta:2005rm}
  A.~Bacchetta, C.~J.~Bomhof, P.~J.~Mulders and F.~Pijlman,
  %``Single spin asymmetries in hadron hadron collisions,''
  Phys.\ Rev.\  D {\bf 72}, 034030 (2005)
  [arXiv:hep-ph/0505268].

\bibitem{metz}
  A.~Metz,
  %``Gluon-exchange in spin-dependent fragmentation,''
  Phys.\ Lett.\  B {\bf 549}, 139 (2002);
  J.~C.~Collins and A.~Metz,
  %``Universality of soft and collinear factors in hard-scattering
  %factorization,''
  Phys.\ Rev.\ Lett.\  {\bf 93}, 252001 (2004)
  [arXiv:hep-ph/0408249];
  S.~Meissner and A.~Metz,
  %``Partonic pole matrix elements for fragmentation,''
  Phys.\ Rev.\ Lett.\  {\bf 102}, 172003 (2009)
  [arXiv:0812.3783 [hep-ph]].

\bibitem{gamberg}
  L.~P.~Gamberg, A.~Mukherjee and P.~J.~Mulders,
  %``Spectral analysis of gluonic pole matrix elements for fragmentation,''
  Phys.\ Rev.\  D {\bf 77}, 114026 (2008).
  [arXiv:0803.2632 [hep-ph]];


\bibitem{Yuan:2009dw}
  F.~Yuan and J.~Zhou,
  %``Collins Fragmentation and the Single Transverse Spin Asymmetry,''
  Phys.\ Rev.\ Lett.\  {\bf 103}, 052001 (2009)
  [arXiv:0903.4680 [hep-ph]].

\bibitem{Yuan:2007nd}
  F.~Yuan,
  %``Azimuthal Asymmetric Distribution of Hadrons Inside a Jet at Hadron
  %Collider,''
  Phys.\ Rev.\ Lett.\  {\bf 100}, 032003 (2008)
  [arXiv:0709.3272 [hep-ph]];
  %``Collins Asymmetry at Hadron Colliders,''
  Phys.\ Rev.\  D {\bf 77}, 074019 (2008)
  [arXiv:0801.3441 [hep-ph]].

\bibitem{trans}
  J.~P.~Ralston and D.~E.~Soper,
  %``Production Of Dimuons From High-Energy Polarized Proton Proton
  %Collisions,''
  Nucl.\ Phys.\  B {\bf 152}, 109 (1979);
  R.~L.~Jaffe and X.~D.~Ji,
  %``Chiral odd parton distributions and polarized Drell-Yan,''
  Phys.\ Rev.\ Lett.\  {\bf 67}, 552 (1991);
  %``Chiral Odd Parton Distributions And Drell-Yan Processes,''
  Nucl.\ Phys.\  B {\bf 375}, 527 (1992).

\bibitem{boer}
  D.~Boer,
  %``Angular dependences in inclusive two-hadron production at BELLE,''
  Nucl.\ Phys.\  B {\bf 806}, 23 (2009)
  [arXiv:0804.2408 [hep-ph]].
  %%CITATION = NUPHA,B806,23;%%


\bibitem{Anselmino:2007fs}
  M.~Anselmino, M.~Boglione, U.~D'Alesio, A.~Kotzinian, F.~Murgia, A.~Prokudin and C.~Turk,
  %``Transversity and Collins functions from SIDIS and e+ e- data,''
  Phys.\ Rev.\  D {\bf 75}, 054032 (2007)
  [arXiv:hep-ph/0701006].

\bibitem{Mulders:1995dh}
  P.~J.~Mulders and R.~D.~Tangerman,
  %``The complete tree-level result up to order 1/Q for polarized
  %deep-inelastic leptoproduction,''
  Nucl.\ Phys.\  B {\bf 461}, 197 (1996)
  [Erratum-ibid.\  B {\bf 484}, 538 (1997)]
  [arXiv:hep-ph/9510301].

\bibitem{Boer:1997nt}
  D.~Boer and P.~J.~Mulders,
  %``Time-reversal odd distribution functions in leptoproduction,''
  Phys.\ Rev.\  D {\bf 57}, 5780 (1998)
  [arXiv:hep-ph/9711485].

\bibitem{Boer:2002ju}
  D.~Boer, S.~J.~Brodsky and D.~S.~Hwang,
  %``Initial state interactions in the unpolarized Drell-Yan process,''
  Phys.\ Rev.\  D {\bf 67}, 054003 (2003)
  [arXiv:hep-ph/0211110].

\bibitem{Boer:2007nd}
  D.~Boer, P.~J.~Mulders and C.~Pisano,
  %``$T^-$ odd effects in photon-jet production at the Tevatron,''
  Phys.\ Lett.\  B {\bf 660}, 360 (2008)
  [arXiv:0712.0777 [hep-ph]];
  %``Dijet imbalance in hadronic collisions,''
  arXiv:0909.4652 [hep-ph].

\bibitem{Lu:2008qu}
  Z.~Lu and I.~Schmidt,
  %``Azimuthal angle dependence of di-jet production in unpolarized hadron
  %scattering,''
  Phys.\ Rev.\  D {\bf 78}, 034041 (2008)
  [arXiv:0805.4006 [hep-ph]].

\bibitem{Bacchetta:2006tn}
  A.~Bacchetta, M.~Diehl, K.~Goeke, A.~Metz, P.~J.~Mulders and M.~Schlegel,
  %``Semi-inclusive deep inelastic scattering at small transverse momentum,''
  JHEP {\bf 0702}, 093 (2007)
  [arXiv:hep-ph/0611265].

\bibitem{Owens:1986mp}
  J.~F.~Owens,
  %``Large Momentum Transfer Production Of Direct Photons, Jets, And
  %Particles,''
  Rev.\ Mod.\ Phys.\  {\bf 59}, 465 (1987).

\bibitem{Lai:1999wy}
  H.~L.~Lai {\it et al.}  [CTEQ Collaboration],
  %``Global QCD analysis of parton structure of the nucleon: CTEQ5 parton
  %distributions,''
  Eur.\ Phys.\ J.\  C {\bf 12}, 375 (2000)
  [arXiv:hep-ph/9903282].

\bibitem{Kretzer:2000yf}
  S.~Kretzer,
  %``Fragmentation functions from flavour-inclusive and flavour-tagged e+ e-
  %annihilations,''
  Phys.\ Rev.\  D {\bf 62}, 054001 (2000)
  [arXiv:hep-ph/0003177].

\end{thebibliography}
\end{document}